\documentclass[prresearch,twocolumn,superscriptaddress]{revtex4-1}
\usepackage{graphicx,amsmath,amssymb}
\usepackage{color}
\usepackage[colorlinks=true, citecolor=blue, urlcolor=blue,citecolor=blue]{hyperref}
\begin{document}
\title{Signatures of quantized coupling between quantum emitters and localized surface plasmons}
\author{Chun-Jie Yang}
\affiliation{School of Physics, Henan Normal University, Xinxiang 453007, China}
\author{Jun-Hong An}\email{anjhong@lzu.edu.cn}
\affiliation{School of Physical Science and Technology, Lanzhou University, Lanzhou 730000, China}
\author{Hai-Qing Lin}\email{haiqing0@csrc.ac.cn}
\affiliation{Beijing Computational Science Research Center, Beijing 100193, China}
\begin{abstract}
Confining light to scales beyond the diffraction limit, quantum plasmonics supplies an ideal platform to explore strong light-matter couplings. The light-induced localized surface plasmons (LSPs) on the metal-dielectric interface acting as a quantum bus have wide potential in quantum information processing; however, the loss nature of light in the metal hinders their application. Here we propose a mechanism to make the reversible energy exchange and the multipartite quantum correlation of a collective of quantum emitters (QEs) mediated by the LSPs persistent. Via investigating the quantized interaction between the QEs and the LSPs supported by a spherical metal nanoparticle, we find that the diverse signatures of the quantized QE-LSP coupling in the steady state, including the complete decay, population trapping, and persistent oscillation, are essentially determined by the different number of bound states formed in the energy spectrum of the QE-LSP system. Enriching our understanding on the light-matter interactions in a lossy medium, our result is instructive in the design of quantum devices using plasmonic nanostructures.
\end{abstract}
\maketitle
\section{Introduction}
Hybrid systems composed of metal nanoparticles (MNPs) and quantum emitters (QEs) have drawn intense attention in physics, chemistry, and materials and life sciences \cite{Giannini2011,Atwater2010,Kabashin2009,Tame2013,WangJF}. By confining light within regions far below the diffraction limit in modes of localized surface plasmons (LSPs), the strong light-matter interaction is realizable in the vicinity of the MNPs \cite{PhysRevB.77.115403,doi:10.1021/acs.nanolett.6b04659,doi:10.1021/acsphotonics.7b00674,Chikkaraddy2016,Savasta2010,Santhosh2016,Chikkaraddy2016,Matsuzak2017,Kewes2018,doi:10.1021/acsphotonics.7b00674,Vasa2018}. Recently, dramatic progress has been made to reveal the modified radiative properties of QEs by the LSPs in quantum plasmonics. Fascinating effects, including the superradiance of an ensemble of dipoles \cite{PhysRevLett.102.077401}, the surface plasmon amplification by stimulated emission of radiation \cite{Noginov2009}, the quantum statistics control of photons \cite{PhysRevLett.105.263601}, and the suppression of quantum fluctuations of light \cite{PhysRevLett.113.263605}, have been found. These effects have led to a wide application of the LSPs in quantum information processing and quantum device designing. However, the dissipation of the LSPs induced by the loss nature of light in metal severely restricts their practical applications \cite{Tame2013,doi:10.1021/acsphotonics.7b01139}.

It has been found that a QE residing near the metal is quenched by its decay through the nonradiative electromagnetic modes absorbed by the metal \cite{PhysRevLett.89.203002,PhysRevLett.96.113002,PhysRevLett.97.017402,Gurlek2018}. Such quenching hampers the complete quantum control in plasmonic systems, where a persistent quantum coherence is of importance \cite{Gurlek2018,PhysRevLett.119.233901}. In the systems of a collective of QEs, the cooperative effect makes the strong coupling between the QEs and the radiative mode dominate the metal absorption \cite{PhysRevLett.112.253601} and suppresses quenching to the QEs \cite{Kongsuwan2018}. It endows the multiple-QE system coupled to metal nanostructures with a promising route to suppress the loss of LSPs in metal \cite{Saez-Blazquez:17,PhysRevA.98.013839}. Going beyond the weak-coupling description of QE-LSP interactions \cite{PhysRevA.94.013854,PhysRevB.95.075412,PhysRevB.97.115402}, it has been found that the LSPs can act as a quantum bus to mediate the coherent interactions and generate the entanglement among QEs \cite{PhysRevB.92.045410,PhysRevA.93.022320,PhysRevA.96.012339}. However, such quantum coherence is dynamically transient and tends to vanish in the long-time limit. In terms of practical applications, persistent quantum coherence and entanglement of the QEs are desired. On the other hand, a widely used description of strong QE-LSP coupling is based on the pseudomode method \cite{PhysRevA.82.043845,Delga_2014,PhysRevLett.119.233901,PhysRevLett.112.253601,PhysRevB.92.205420,PhysRevLett.117.107401}, which decomposes the spectrum into a sum of discrete resonant modes with Lorentzian expansion and succeeds in mapping the non-Markovian dynamics into a Markovian one \cite{PhysRevB.89.041402}. When the coupling is strong enough, the QEs and LSPs are highly hybridized, and thus the pseudomode method is insufficient and a rigorous continuous-mode theory is needed.

\begin{figure}[tbp]
\includegraphics[width=0.4\columnwidth]{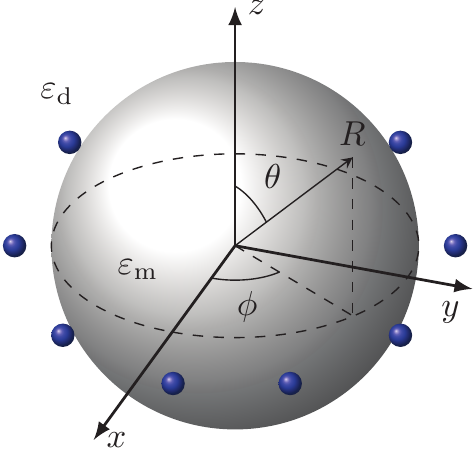}
\caption{Schematic diagram of $N$ QEs positioned at $\mathbf{r}_{l}$ on the equator plane of the MNP with radius $R$ and permittivity $\varepsilon_{\text{m}}(\omega)$. The system is put in a homogeneous and isotropic medium with dielectric constant $\varepsilon_\text{d}$. } \label{Fig1}
\end{figure}

In this paper, going beyond the pseudomode method, we study exactly the dissipative dynamics of a collective of QEs interacting with the LSPs supported by a MNP. A mechanism to overcome the loss effect of the LSPs in the metal is discovered. We find the diverse signatures of the strong QE-LSP couplings, including complete decay, population trapping, and persistent oscillations, in the long-time steady state. Our analyses reveal that they are determined by the formation of different numbers of QE-LSP bound states. We also find that, as a consequence of the suppression of loss effect of the LSPs, a persistent entanglement among the QEs can be mediated by the LSPs. Such bound-state-favored persistent entanglement among the QEs plays a constructive role in applying the LSPs as a quantum bus in quantum information processing.


\section{System and quantization}\label{sys}
The system is composed of a MNP surrounded by $N$ QEs. The QEs labeled by $l$ are positioned at $\mathbf{r}_{l}$ on the equator plane of the MNP (see Fig. \ref{Fig1}). Each QE is modeled as a two-level system with frequency $\omega_{l}$ and dipole moment $\pmb{\mu}_{l}$. The MNP has a radius $R$ and a dielectric permittivity denoted by a complex Drude model $\varepsilon_\text{m} (\omega )=\varepsilon _{\infty }-\omega _{p}^{2}/[\omega (\omega +i\gamma _{p})]$, where $\omega _{p}$ is the bulk plasma frequency, $\varepsilon _{\infty}$ is the high-frequency limit of $\varepsilon_{\text{m}}(\omega)$, and $\gamma _{p}$ is the Ohmic loss of light in the MNP \cite{PhysRevB.6.4370}. The whole system is embedded in a homogeneous medium with dielectric constant $\varepsilon_\text{d}$. We consider that both of the dielectric and the metal are nonmagnetic and thus their permeability $\mu_\text{d}=\mu_\text{m}\equiv1$.

Besides propagating into the dielectric as a radiative mode and being absorbed by the MNP as a nonradiative mode, the optical field emitted by the QE also induces a confined hybrid mode which consists of LSPs localized near the metal-dielectric interface \cite{Pitarke2007}. The LSPs enable a confinement of light within the subwavelength areas on the interface, which supplies an ideal platform to explore the strong quantized light-matter coupling \cite{PhysRevB.89.041402,PhysRevB.95.161408}. A quantization method of light in the absorbing medium has been proposed based on the dyadic Green's function, where the absorption of the medium to light is described by a Langevin noise \cite{PhysRevA.53.1818,PhysRevA.57.3931}. Then the electric field reads
\begin{equation}
\hat{\mathbf{E}}(\mathbf{r},\omega )=\frac{ic^{-2}\omega ^{2}}{\sqrt{\pi
\varepsilon _{0}/\hbar }}\int d^{3}\mathbf{r}'\sqrt{\text{Im}[\varepsilon _\text{m}(\omega )]}\mathbf{G}(\mathbf{r},\mathbf{r}',\omega )\cdot \mathbf{\hat{f}}(\mathbf{r}',\omega ), \nonumber
\end{equation}
where $\varepsilon_{0}$ is the vacuum permittivity, $c$ is the speed of light, and $\hat{\bf{f}}(\bf{r},\omega)$ satisfying $[\hat{\bf{f}}(\bf{r},\omega),\hat{\bf{f}}^\dag(\bf{r}',\omega')]=\delta(\bf{r}-\bf{r}')\delta({\omega-\omega'})$ is the annihilation operator of light. The Green's function $\bf{G(\bf{r},\bf{r}^{\prime},\omega)}$ satisfying the Helmholtz equation $[{\pmb\nabla}\times{\pmb\nabla}\times-\omega^{2}c^{-2}\varepsilon_\text{m}(\omega)]\mathbf{G}(\mathbf{r},\mathbf{r}^{\prime },\omega )=\mathbf{I}\delta (\mathbf{r}-\mathbf{r}^{\prime })$, with $\mathbf{I}$ being the identity matrix, denotes the field in frequency $\omega$ evaluated at $\bf{r}$ due to a point source at $\bf{r}^{\prime}$. The spatial distribution of all of the three modes has been incorporated in $\bf{G(\bf{r},\bf{r}^{\prime},\omega)}$ by solving the Helmholtz equation subject to the boundary condition of the system geometry. It allows for a complete description of the quantized light-matter coupling by calculating $\bf{G(\bf{r},\bf{r}^{\prime},\omega)}$. For a spherical MNP, the Green's function is analytically solvable. For more details see Appendix \ref{appen-Green}.

The Hamiltonian of the full QE-MNP system under the dipole and rotating-wave approximations reads \cite{PhysRevA.62.053804}
\begin{eqnarray}\label{Hmt}
\hat{H} &=&\sum_{l=0}^{N-1}\hbar \omega _{l}\hat{\sigma}^{\dag}_{l}\hat{\sigma}_{l}+\int d^{3}\mathbf{r}\int d\omega \hbar \omega \hat{\mathbf{f}}^{\dag }(\mathbf{r},\omega )\cdot \hat{\mathbf{f}}(\mathbf{r},\omega )\nonumber\\
&&-\sum_{l=0}^{N-1}\int d\omega \lbrack \mathbf{\pmb\mu }_{l}\cdot \hat{\mathbf{E}}(\mathbf{r}_{l},\omega )\hat{\sigma}^{\dag}_{l}+\text{H.c.}],
\end{eqnarray}
where $\hat{\sigma}_{l}=|g_{l}\rangle \langle e_{l}|$ is the transition operator from the excited state $|e_{l}\rangle$ to the ground state $|g_{l}\rangle$ of the $l$th QE. The validity of the rotating-wave approximation in a related system was revealed in \cite{PhysRevB.97.115402}. The dipole approximation works when the QE size is sufficiently small \cite{PhysRevB.82.115334,PhysRevB.86.085304,PhysRevA.93.053803}. Conventionally, the LSPs are viewed as a few discrete pseudomodes with Lorentzian expansion. Then one can use the standard cavity QED method to describe the QE-LSP coupling \cite{PhysRevA.82.043845,Delga_2014,PhysRevLett.119.233901,PhysRevLett.112.253601,PhysRevB.92.205420,PhysRevLett.117.107401}. It neglects the non-Lorentzian features of the spectrum and may be insufficient when the QE is close to the interface \cite{PhysRevB.95.161408}, where the hybridization of the QEs and the LSPs dominates.
\section{Exact dynamics}\label{dyn}
We can see that the total excitation number $\hat{\mathcal{N}}=\sum_l\hat{\sigma}^{\dag}_{l}\hat{\sigma}_{l}+\int d^{3}\mathbf{r}\int d\omega \mathbf{\hat{f}}^{\dag }(\mathbf{r},\omega )\cdot \mathbf{\hat{f}}(\mathbf{r},\omega )$ is conserved. In the single-excitation subspace, the time-evolved state can be expanded as $
|\Psi(t)\rangle=[\sum _lc_{l}(t)\hat{\sigma}_l^\dag+\int d^{3}\mathbf{r}\int d\omega d_{\mathbf{r},\omega}(t) \hat{\mathbf{f}}^\dag(\mathbf{r},\omega)]|G;\{0_{\omega}\}\rangle$,
where $|G\rangle$ denotes all the QEs in the ground state and $|\{0_{\omega}\}\rangle $ is the vacuum state of the total modes. It can be derived that $c_{l}(t)$ obeys (see Appendix \ref{appen-evolution})
\begin{equation}
\mathbf{\dot{c}}(t)+i\omega _{0}\mathbf{c}(t)+\int_{0}^{t}d\tau
\int_{0}^{\infty }d\omega e^{-i\omega (t-\tau )}\mathbf{J}(\omega )\mathbf{c}
(\tau )=0, \label{evolution}
\end{equation}
where $\mathbf{c}(t)=(c_{0}(t),\ldots,c_{N-1}(t))^{T}$ is a column vector with $c_l(t)$ being the excited-state probability amplitude of $l$th QE, and $\mathbf{J}(\omega)$ is a matrix, with $J_{lj }(\omega )=\omega ^{2}\pmb{\mu}_{l}\cdot\textrm{Im}[ \mathbf{G}(\mathbf{r}_{l}, \mathbf{r}_{j},\omega )]\cdot\pmb{\mu}_{j}^{\ast }/(\pi\hbar \varepsilon_{0}c^{2})$ the correlated spectral densities between the $l$th and $j$th QEs. Thus all the actions of the metal-dielectric structure on the QEs have been collected in $\mathbf{J}(\omega)$. We have chosen the QEs having identical frequency $\omega_l=\omega_{0}$ and used $\int d^3\mathbf{s}\frac{\omega ^{2}}{c^{2}}$Im$[\varepsilon _{m}(\omega )]\mathbf{G}(\mathbf{r},\mathbf{s},\omega )\mathbf{G}^{\ast }(\mathbf{r}^{\prime },\mathbf{s},\omega )=\textrm{Im}[\mathbf{G}(\mathbf{r},\mathbf{r}^{\prime },\omega )]$ \cite{PhysRevA.62.053804}. The convolution in Eq. \eqref{evolution} renders the QE dynamics non-Markovian. The correlation of different $c_l(t)$ indicates that, although direct couplings of QEs in Eq. (\ref{Hmt}) are absent, their indirect couplings can be effectively induced by exchanging the virtual excitations of the photons.

The solution of Eq. \eqref{evolution} can be analyzed by a Laplace transform, which yields $\tilde{\mathbf{c}}(s)=\mathbf{V}\bar{\mathbf{c}}(s)\mathbf{V}^{-1}\mathbf{c}(0)$, with $\bar{\mathbf{c}}(s)=[s+i\omega _{0}+\int_{0}^{\infty }d\omega \frac{\mathbf{D}(\omega )}{s+i\omega }]^{-1}$. We have used the Jordan decomposition of $\mathbf{J}(\omega )=\mathbf{V}\mathbf{D}(\omega )\mathbf{V}^{-1}$, with $\mathbf{V}$ and $\mathbf{D}(\omega )=\text{diag}[D_{0}(\omega),\ldots,D_{N-1}(\omega)]$ its similarity matrix and Jordan canonical form, respectively. Then $\mathbf{c}(t)$ is obtainable by in inverse Laplace transform to $\bar{\mathbf{c}}(s)$, which can be done by finding its poles from
\begin{equation}\label{bound}
y_{l}(\varpi )\equiv\omega _{0}-\int_{0}^{\infty }\frac{D_{l}(\omega )}{\omega -\varpi }d\omega =\varpi, ~~\varpi =is.
\end{equation}
It can be proven that the roots $\varpi $ multiplied by $\hbar $ are just the hybrid eigenenergies of the QEs and the LSPs in the single-excitation subspace (see Appendix \ref{appen-eigenenergy}). Since $y_{l}(\varpi )$ is a monotonically decreasing function when $\varpi <0$, each one of Eqs. (\ref{bound}) has one discrete root $\varpi^{b} _{l}$ if $y_{l}(0)<0$. We call the discrete eigenstates with eigenenergy $\hbar \varpi _{l}^{b}$ the bound state. In the region $\varpi >0$, it has an infinite number of roots, which form a continuous energy band. Determined by the system parameters, at most $N$ independent bound states could be formed. Using Cauchy's residue theorem, we readily have $\mathbf{c}(t)=\mathbf{V}\bar{\mathbf{c}}(t)\mathbf{V}^{-1}\mathbf{c}(0)$ with the elements of $\bar{\mathbf{c}}(t)$ as \cite{PhysRevB.95.161408}
\begin{equation}
\bar{c}_{l}(t)=Z_{l}e^{-i\varpi _{l}^{b}t}+\int_{i\epsilon +0}^{i\epsilon+\infty }\frac{d\varpi }{2\pi }\bar{c}_{l}(-i\varpi )e^{-i\varpi t},\label{barc}
\end{equation}
where the first term with $Z_{l}=[1+\int_{0}^{\infty }\frac{D_{l}(\omega )}{(\varpi _{l}^{b}-\omega )^{2}}d\omega ]^{-1}$ is from the bound state and the second term is from the energy band. Oscillating with time in continuously changing frequencies, the second term behaves as a decay and
tends to zero due to out-of-phase interference. Thus, if the bound state is absent, then $\lim_{t\rightarrow \infty }\mathbf{c}(t)=0$ characterizes a complete decay; while if the bound states are formed, then $\lim_{t\rightarrow \infty }\mathbf{c}(t)=\mathbf{V(Z}e^{-i\pmb{\varpi }^{b}t}\mathbf{)V}^{-1}\mathbf{c}(0)$, with $\bf{x}=\text{diag}(x_{0},\ldots,x_{N-1})$ for $\bf{x}=\bf{Z}$ and $\pmb{\varpi}^b$, implies decoherence suppression. This indicates that the dynamics of the QEs in the long-time limit is intrinsically determined by the energy-spectrum characters of the whole QE-LSP system. Generally, solving $\mathbf{V}$ and $\mathbf{D}(\omega)$ needs numerical calculations. Here, for concreteness, we choose that all the QEs have identical dipole moments and uniform coordinates $\mathbf{r}_{l}=(r,\pi/2,2\pi l/N)$ such that $\mathbf{J}(\omega )$ is a symmetric circulant matrix with $J_{lj}(\omega)=J_{mn}(\omega)\equiv J_{|l-j|}(\omega) $ for $|l-j|=|m-n|$ (see Appendix \ref{appen-Green}). Because $\mathbf{J}(\omega)$ is a symmetric circulant matrix, we readily have $D_{l}(\omega )=\sum_{j=0}^{N-1}J_{j}(\omega )\lambda_{l}^{N-j}$ and $\mathbf{V}=(\upsilon _{0},\ldots ,\upsilon _{N-1})$ with $\upsilon _{l}=\frac{1}{\sqrt{N}}(1,\lambda  _{l},\ldots ,\lambda _{l}^{N-1})^{T}$ and $\lambda _{l}=\exp (-2\pi il/N)$ \cite{CIT-006}.

\section{Results and discussion}
\begin{figure}[tbp]
\centering
\includegraphics[width=\columnwidth]{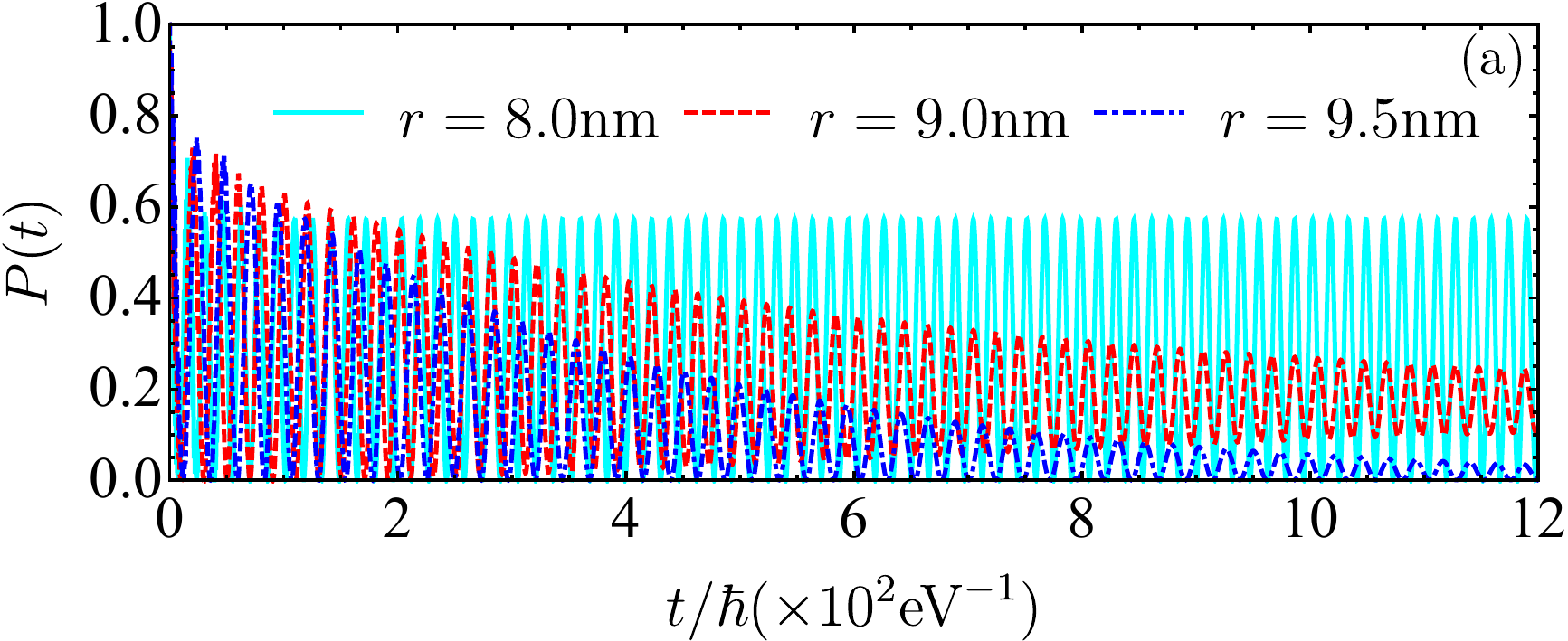}\\
\includegraphics[width=.95\columnwidth]{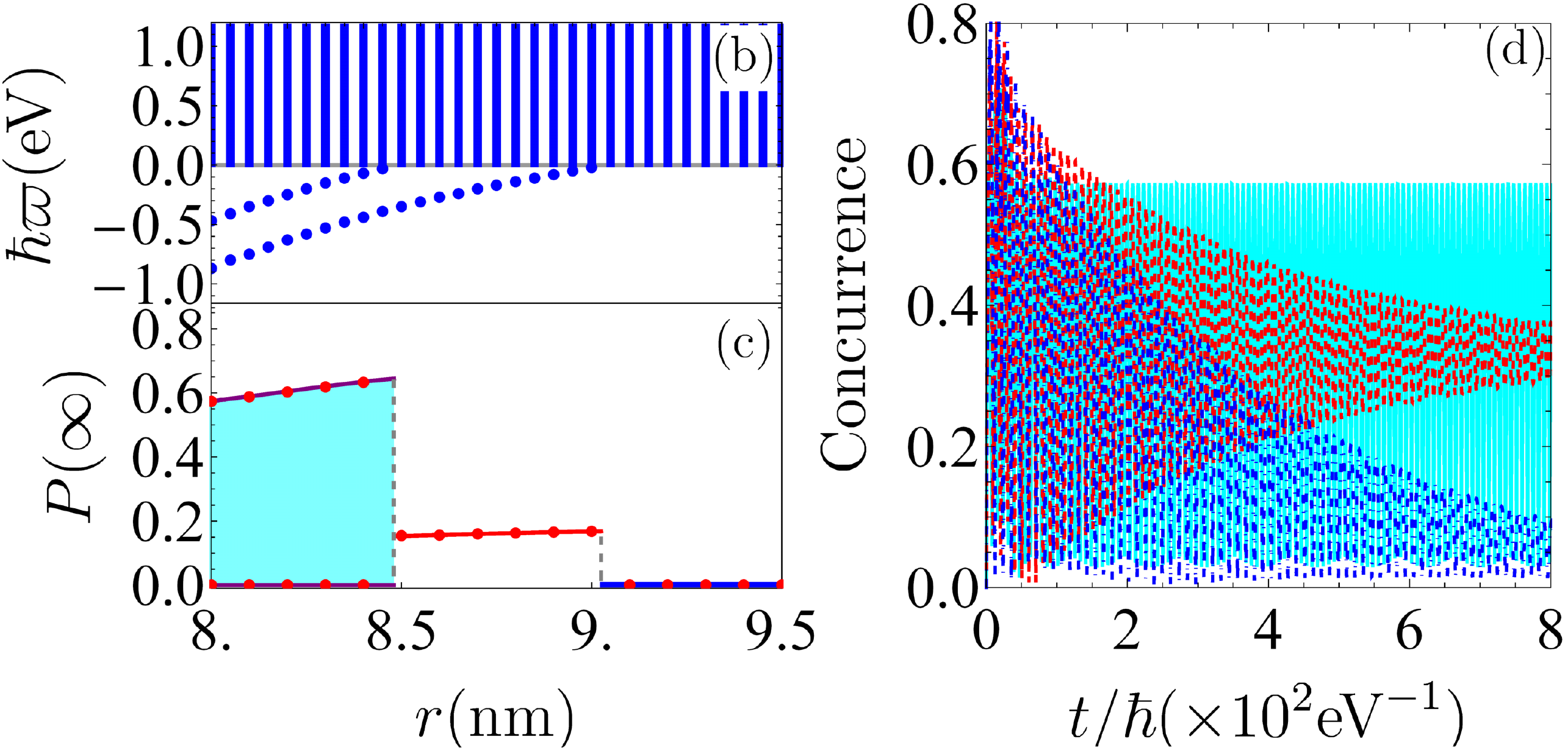}\\
\caption{(a) Evolution of $P(t)$ in different $r$ obtained by numerically solving Eqs. \eqref{evolution}. (b) Energy spectrum of the whole system in different $r$. Two branches of bound states are formed in the band gap. (c) Long-time values of $P(t)$ obtained from the exact dynamics (red dots) and from Eq. \eqref{asympt} (solid lines). The cyan region covers the values of $P(\infty)$ during its persistent oscillation. (d) Evolution of concurrence obtained by solving Eqs. \eqref{evolution}. The other parameters are $N=2$, $\hbar\omega_{0}=0.8$ eV, and $R=5$ nm. } \label{Fig2}
\end{figure}
It was previously found that the reversible energy exchange between the QEs induced by a common surface plasma tends to vanish in the long-time limit under the Born-Markovian approximation \cite{PhysRevLett.106.020501}. Different from that result, we will show that such mediated coherent coupling can induce a persistently reversible energy exchange between the QEs even in the steady state when the approximation is relaxed. We choose silver for the metal with $\hbar\omega _{p}=9.01$ eV, $\varepsilon _{\infty}=3.718$, and $\hbar\gamma _{p}=0.09$ eV in the interested frequency range \cite{Scholl2012} and the QEs with $\hbar \gamma_{0}=0.1$ meV.
We focus on the QE dynamics by studying the initial-state fidelity $P(t)=|\langle \Psi (0)|\Psi (t)\rangle |^{2}$.

First, taking $N=2$, we consider that only one of the QEs is excited initially, i.e., $|\Psi (0)\rangle =\hat{\sigma}_0^{\dag}|G;\{0_\omega\}\rangle$. We can calculate that with time evolution the fidelity reads $P(t)=|c_0(t)|^2$. Figure \ref{Fig2}(a) shows the evolution of $P(t)$ in three characteristic values of $r$. As a result of the near-field enhancement of the LSPs, a significant oscillation appears in the dynamics for all three cases. Absent in the Born-Markovian approximate result, this is entirely the non-Markovian effect, which represents a reversible energy exchange and thus manifests the strong coupling between the QEs mediated by the LSPs \cite{PhysRevB.89.041402}. It is interesting to see that the non-Markovian effect manifests its action on the QEs not only in its transient dynamics, but also in its steady state. When $r=9.5$ nm, $P(t)$ tends to zero accompanying the QEs decay completely to the ground state, which is consistent with the previous results \cite{He2012,PhysRevB.92.125432}. However, a remarkable difference appears with further decreasing $r$. One can see that $P(t)$ tends to a nonzero value when $r=9.0$ nm, which represents a stable population trapping in the system, while when $r=8.0$ nm, $P(t)$ tends to a lossless oscillation with a constant frequency, which is quite like the Rabi oscillation \cite{dudin2012observation} and represents a persistent energy exchange among QEs caused by the QE-LSP interaction. These diverse signatures can be explained by our bound-state analysis. From Eq. \eqref{barc} we have (see Appendix \ref{appen-solution})
\begin{equation} \label{asympt}
\lim_{t\rightarrow\infty}|P(t)|^2=\left\{ \begin{aligned}
         &0,\hspace{2.5cm}~~~M=0\\
         &Z^2/4,\hspace{1.9cm}~~~M=1 \\
         &[Z_0^2+Z_1^2+D(t)]/4,~M=2,
                          \end{aligned} \right.
\end{equation}where $M$ is the number of formed bound states and $D(t)=2Z_0Z_1\cos[(\varpi_1^b-\varpi_0^b)t]$ is the interference between the two bound states. This conclusion can be confirmed by the energy spectrum shown in Fig. \ref{Fig2}(b). The two branches of bound states formed in the band gap divide the spectrum into three regions: without bound state when $r\gtrsim9.0$ nm, one bound state when $8.5\lesssim r\lesssim9.0$ nm, and two bound states when $r\lesssim8.5$ nm. The regions match well with the ones where $P(\infty)$ shows different behaviors [see Fig. \ref{Fig2}(c)], i.e., complete decay, population trapping, and persistent oscillation, as expected from Eq. (\ref{asympt}). Such bound-state-favored behaviors are constructive to generate entanglement between the QEs. Different from the asymptotic vanishing in the Born-Markovian approximation \cite{PhysRevLett.106.020501} and in the absence of the bound state, the generated entanglement can be preserved as long as the bound states are formed [see Fig. \ref{Fig2}(d)]. This is helpful for utilizing plasmonic nanostructures in designing quantum devices. Our results can be generalized to the case of a large number of QEs. With more of the bound states being formed in the large-$N$ case, the persistent oscillations will be complicated, but the mechanism is the same as in the present case. In Appendix \ref{appen-4QEs}, the dynamics for $N=4$ is provided. Note that the similar bound-state-induced decoherence suppression for the single-QE case has been found in Refs. \cite{PhysRevB.95.075412,PhysRevB.95.161408}.
\begin{figure}[tbp]
\centering
\includegraphics[width=.96\columnwidth]{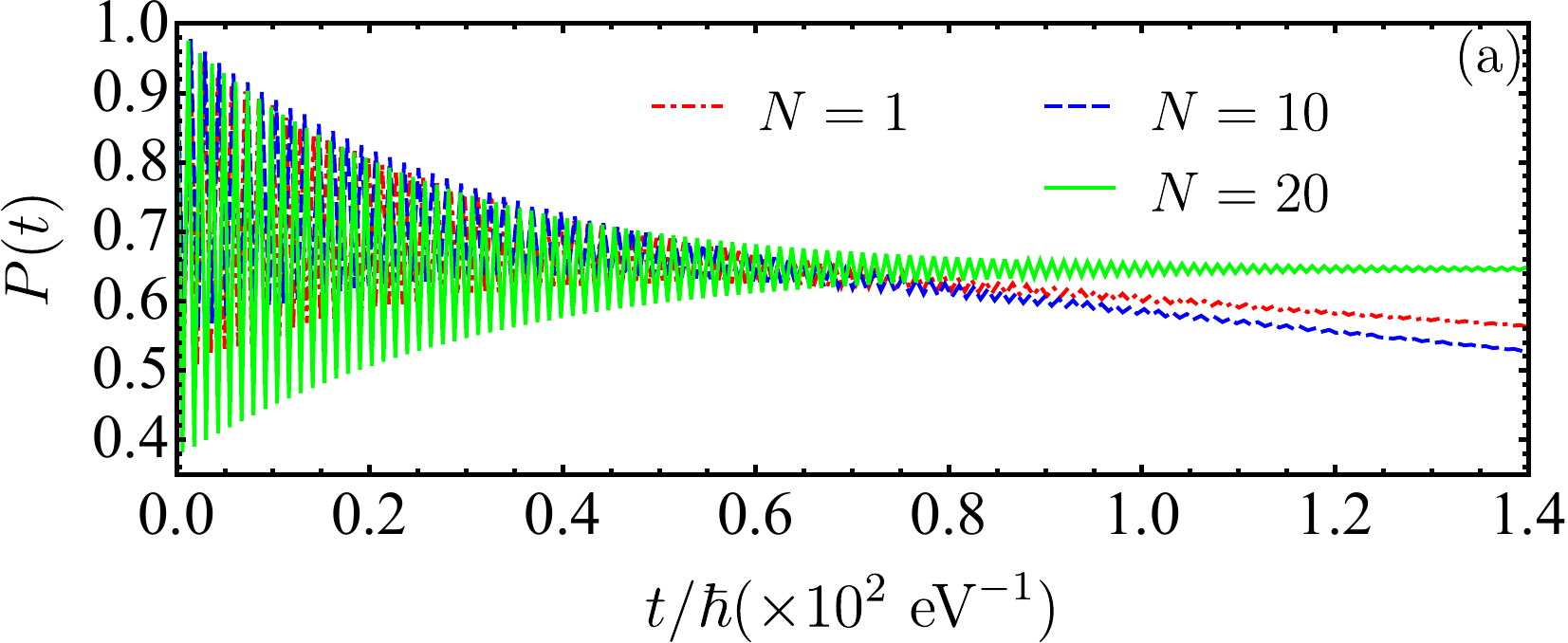}\\
\includegraphics[width=\columnwidth]{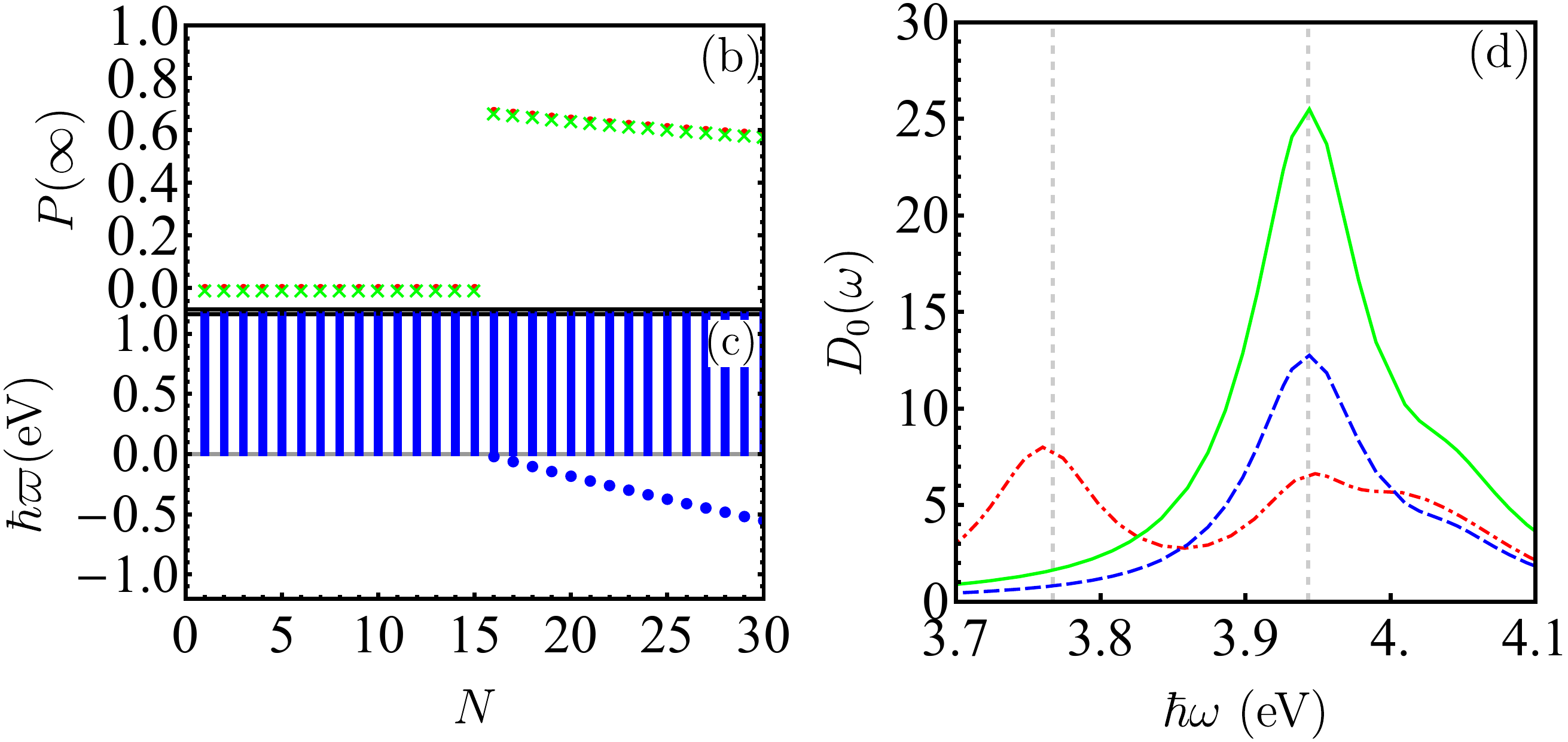}\\
\caption{(a) Evolution of $P(t)$ with $r=9.5$ nm in different $N$ obtained by the exact dynamics. (b) Long-time values of $P(t)$ obtained by the exact dynamics (red dots) and the bound-state analysis (green crosses). (c) Eigenenergy in different $N$. (d) Spectral density $D_{0}{(\omega)}$ and frequencies of the dipole and quadrupole modes of the LSPs (gray dashed lines). The other parameters are the same as in Fig. \ref{Fig2}.} \label{Fig3}
\end{figure}

Next we consider that the QEs are initially in a $|\Psi(0)\rangle =\frac{1}{\sqrt{N}}\sum_{l =0}^{N-1}\hat{\sigma}^{\dag}_{l}|G;\{0_\omega\}\rangle$, which is a multipartite entangled state widely used in quantum information processing \cite{1367-2630-5-1-136,GORBACHEV2003267}. The canonical transformation $\mathbf{V}$ can convert Eq. \eqref{evolution} into $\dot{\bar{\mathbf{c}}}(t)+i\omega _{0}\bar{\mathbf{c}}(t)+\int_{0}^{t}d\tau \int d\omega e^{-i\omega (t-\tau )}\mathbf{D}(\omega )\bar{\mathbf{c}}(\tau)=0$ with $\bar{\mathbf{c}}(t)\equiv\mathbf{V}^{-1}\mathbf{c}(t)$. Its initial condition can be calculated as $\bar{\mathbf{c}}(0)=(1,0,\ldots,0)$, under which only the $\bar{c}_0(t)$ component of this matrix equation has a nonzero solution. Thus its dynamics has the same equation of motion as the one of a single QE coupled to the LSPs \cite{PhysRevB.95.161408}. This indicates that the $N$ QEs collectively act as a two-level superatom to interact with the LSPs with the spectral density characterized by $D_{0}(\omega)$. This notion of a superatom is a powerful concept in designing single-photon quantum sources \cite{RevModPhys.82.2313,PhysRevX.7.041010,PhysRevX.5.031015,PhysRevLett.121.013601}. We can calculate the initial-state fidelity $P(t)=|\bar{c}_{0}(t)|^2$.

In the same mechanism as the case of $N=2$, the entanglement of the QEs can be preserved in the steady state due to the formation of the bound state. Figure \ref{Fig3}(a) shows the evolution of $P(t)$ for a different number $N$ of QEs. It shows that $P(t)$ tends to a finite value for large $N$, where the QEs remain entangled. It can be understood from the bound-state analysis. As discussed above, we readily obtain $\lim_{t\rightarrow \infty }\bar{c}_{0}(t)=Z_{0}e^{-i\varpi_{0}^{b}t}$ when Eq. \eqref{bound}, with $l=0$, has an isolate root in the region $\varpi <0$. Figures \ref{Fig3}(b) and \ref{Fig3}(c) show that the region where $P(t)$ tends to a stable value matches well with the one where a bound state is formed in the energy spectrum of the whole system. This verifies again our conclusion that it is the formation of a bound state that preserves the entanglement in the steady state.  We also plot in Fig. \ref{Fig3}(d) the spectral density $D_{0}(\omega)$, which measures the coupling strength of the QEs and the LSPs. We can see that the contribution of the resonant dipole mode $\omega_1=3.77$ eV is entirely canceled, while the one of the quadrupole mode $\omega_2=3.94$ eV is enhanced by increasing $N$ (see Appendix \ref{appen-Green}). This is due to the destructive interference of the undistinguished coupling channels between different QEs and the LSPs \cite{PhysRevLett.112.253601,PhysRevLett.121.013601,LUG}.

We note that, although we consider only the case that the dipole moments of the QEs are polarized along the radial direction, our result can be generalized to other cases. Some quantitative difference might occur, but the constructive role played by the bound states in overcoming the loss effect of the LSPs in the MNP does not change. We emphasize that our finding is realizable in the state-of-art technique of experiments. The parameters used in our calculation are near the ones of silver as the MNP and the $J$ aggregates as the QEs. Their strong coupling has been studied \cite{Schlather,PhysRevLett.97.266808,FIDDER1990529,PhysRevLett.108.066401}. The bound state and its distinguished role in the non-Markovian dissipative dynamics have recently been observed in both photonic crystal \cite{Liu2017} and ultracold-atom systems \cite{Kri2018}. This means that our finding is completely realizable in quantum plasmonics system, where the strong light-matter coupling is more manifest than in other systems.

\section{Conclusion}\label{con}
We have proposed a mechanism to overcome the loss effect of LSPs in metal by investigating the exact dynamics of $N$ QEs coupled to LSPs supported by a MNP. It has been found that, in sharp contrast to the previous approximate result that the reversible energy exchange and the entanglement of the QEs mediated by the LSPs exclusively tends to vanish due to the loss effect of LSPs in metal, the persistent quantum coherence and entanglement can be established among the QEs by the LSPs. Our analysis indicates that it is the formation of hybrid bound states in the energy spectrum of the QE-LSP system that governs this lossless behavior. Such bound-state-assisted behavior is helpful in the application of LSPs as a quantum bus. The further study of the multipartite $W$-class state demonstrates the collective suppression of the resonant dipole mode and the enhancement of the quadrupole mode in the QE-LSP coupling. Within the present experimental state of the art, our finding supplies a guideline for experiments to design
quantum devices using the plasmonic nanostructures.
\section{Acknowledgments}
The work was supported by the Natural Science Foundation of China (Grants No. 11704103, No. 11875150, and No. 11834005), by the Doctoral Scientific Research Foundation of Henan Normal University (Grant No. 5101029170296), and by the Fundamental Research Funds for the Central Universities of China.
\appendix

\section{Green's function of the spherical metal nanoparticle} \label{appen-Green}
In this appendix, we give the derivations of the Green's function of the spherical MNP in calculating the exact dynamics of QEs coupled to LSPs.

Given a spherical MNP with permittivity $\varepsilon_{\text{m}}(\omega)$ and radius $R$ embedded in a homogeneous medium with dielectric constant $\varepsilon_{\text{d}}$, the Green's functions contributed by the free-space radiation sources and by the MNP-QE interaction are given by \cite{PhysRevA.93.022320,PhysRevB.85.075303,PhysRevA.89.053835}
\begin{widetext}
\begin{eqnarray}
\mathbf{G}^{0}(\mathbf{r},\mathbf{r}^{\prime },\omega ) &=&-\frac{\mathbf{\hat{r}\hat{r}}\delta (\mathbf{r}-\mathbf{r}^{\prime })}{k_{1}^{2}}+\frac{ik_{1}}{4\pi }\sum_{e,o}\sum_{n=1}^{\infty }\sum_{m=0}^{n}(2-\delta
_{0m})\frac{2n+1}{n(n+1)}\frac{(n-m)!}{(n+m)!}\nonumber \\
&&\times\left\{
\begin{array}{c}
\big[\mathbf{M}_{mn_{o}^{e}}^{(1)}(k_{1}\mathbf{r})\mathbf{M}_{mn_{o}^{e}}(k_{1}\mathbf{r}^{\prime })+\mathbf{N}_{mn_{o}^{e}}^{(1)}(k_{1}\mathbf{r})\mathbf{N}_{mn_{o}^{e}}(k_{1}\mathbf{r}^{\prime })\big],\ \mathbf{\hat{r}>\hat{r}}^{\prime }\\
\big[\mathbf{M}_{mn_{o}^{e}}(k_{1}\mathbf{r})\mathbf{M}_{mn_{o}^{e}}^{(1)}(k_{1}\mathbf{r}^{\prime })+\mathbf{N}_{mn_{o}^{e}}(k_{1}\mathbf{r})\mathbf{N}_{mn_{o}^{e}}^{(1)}(k_{1}\mathbf{r}^{\prime })\big],\ \mathbf{\hat{r}<\hat{r}}^{\prime }
\end{array}
\right.\label{dir}\\
\mathbf{G}^{\text{R}}(\mathbf{r},\mathbf{r}^{\prime },\omega )&=&\frac{ik_{1}}{4\pi }\sum_{e,o}\sum_{n=1}^{\infty }\sum_{m=0}^{n}(2-\delta _{0m})\frac{2n+1}{n(n+1)}\frac{(n-m)!}{(n+m)!}\big[\mathcal{R}^{H}\mathbf{M}_{mn_{o}^{e}}^{(1)}(k_{1}\mathbf{r})\mathbf{M}_{mn_{o}^{e}}^{(1)}(k_{1}\mathbf{r}^{\prime })\nonumber\\
&&+\mathcal{R}^{V}\mathbf{N}_{mn_{o}^{e}}^{(1)}(k_{1}\mathbf{r})\mathbf{N}_{mn_{o}^{e}}^{(1)}(k_{1}\mathbf{r}^{\prime })\big],\label{sca}
\end{eqnarray}
\end{widetext}
where $\mathcal{R}^{H}$ and $\mathcal{R}^{V}$ are the scattering coefficients corresponding to the transverse electric field $\mathbf{M}_{mn_{o}^{e}}$ and the transverse magnetic field $\mathbf{N}_{mn_{o}^{e}}$ with even and odd contributions. According to the boundary conditions at the surface, $\mathcal{R}^{H}$ and $\mathcal{R}^{V}$ are given by
\begin{equation} \label{coeff}
\mathcal{R}^{H}=\frac{\tau _{2}\partial \tau _{1}-\tau _{1}\partial \tau _{2}}{\kappa _{1}\partial \tau_{2}-\tau _{2}\partial \kappa _{1}},\ \mathcal{R}^{V}=\frac{k_{1}^{2}\tau _{1}\partial \tau _{2}-k_{2}^{2}\tau _{2}\partial
\tau _{1}}{k_{2}^{2}\tau _{2}\partial \kappa _{1}-k_{1}^{2}\kappa_{1}\partial \tau _{2}},
\end{equation}
where $\tau _{i}=j_{n}(k_{i}R)$, $\kappa _{i}=h_{n}^{(1)}(k_{i}R)$, $\partial \tau_{i}=\partial _{\rho }[\rho j_{n}(\rho )]_{\rho =k_{i}R}$, and $\partial \kappa
_{i}=\partial _{\rho }[\rho h_{n}^{(1)}(\rho )]_{\rho =k_{i}R}$. Here, $j_{n}(x)$ and $h_{n}^{(1)}(x)$ are the spherical Bessel functions and the Hankel functions of the first kind, respectively, with $k_{1}=\omega \sqrt{\varepsilon _{\text{d}}}/c$ and $k_{2}=\omega \sqrt{\varepsilon _{\text{m}}(\omega )}/c$ the wave vectors in the dielectric and the metal. The vector functions in spherical coordinates are defined as
\begin{widetext}
\begin{eqnarray}
\mathbf{M}_{mn}^{e}(k\mathbf{r}) &=&-j_{n}(kr)\Big[\frac{m}{\sin \theta }P_{n}^{m}(\cos \theta )\sin m\varphi \hat{\pmb{\theta}}+\frac{dP_{n}^{m}(\cos \theta )}{d\theta }\cos m\varphi\hat{\pmb{\varphi}}\Big], \\
\mathbf{M}_{mn}^{o}(k\mathbf{r}) &=&j_{n}(kr)\Big[\frac{m}{\sin \theta }P_{n}^{m}(\cos \theta )\cos m\varphi \hat{\pmb{\theta}}-\frac{dP_{n}^{m}(\cos \theta )}{d\theta }\sin m\varphi\hat{\pmb{\varphi}}\Big], \\
\mathbf{N}_{mn}^{e}(k\mathbf{r}) &=&\frac{n(n+1)}{kr}j_{n}(kr)P_{n}^{m}(\cos \theta )\cos m\varphi \hat{\mathbf{r}}+\frac{1}{kr}\frac{d[rj_{n}(kr)]}{dr}\left[ \frac{dP_{n}^{m}(\cos \theta )}{d\theta }\cos m\varphi \hat{\pmb{\theta}}-\frac{m}{\sin \theta }P_{n}^{m}(\cos \theta )\sin m\varphi \hat{\pmb{\varphi}}\right], ~~~~~~ \\
\mathbf{N}_{mn}^{o}(k\mathbf{r}) &=&\frac{n(n+1)}{kr}j_{n}(kr)P_{n}^{m}(\cos \theta )\sin m\varphi \mathbf{\hat{r}+}\frac{1}{kr}\frac{d[rj_{n}(kr)]}{dr}\left[ \frac{dP_{n}^{m}(\cos \theta )}{d\theta }\sin m\varphi \hat{\pmb{\theta}}+\frac{m}{\sin \theta }P_{n}^{m}(\cos \theta )\cos m\varphi \hat{\pmb{\varphi}}\right].~~~
\end{eqnarray}
\end{widetext}
where $P_{n}^{m}(x)$ are the associated Legendre polynomials. In Eqs. \eqref{dir} and \eqref{sca}, the superscript $(1)$ denotes that $j_{n}(x)$ has to be replaced by $h_{n}^{(1)}(x)$.

In the case that the dipole moments of the QEs are polarized along the radial direction, only the $rr$ component of the Green's function contributes to the interactions. In the structure studied, the QEs labeled by $l$ are located at $\mathbf{r}_{l}=(r,\pi/2,2\pi l/N)$, with $l=0,\ldots,N-1$. We obtain
\begin{widetext}
\begin{equation}\label{gzz}
\mathbf{G}_{rr}(\mathbf{r}_{l},\mathbf{r}_{j},\omega )=-\frac{%
\delta (\mathbf{r}_{l}-\mathbf{r}_{j})}{k_{1}^{2}}+\frac{ik_{1}}{%
4\pi }\sum_{n=1}^{\infty }\sum_{m=0}^{n}c_{mn}\cos [\frac{2\pi m(l-j)}{N}]\frac{h_{n}^{(1)}(k_{1}r)[j_{n}(k_{1}r)+\mathcal{R}%
^{V}h_{n}^{(1)}(k_{1}r)]}{[k_{1}r/P_{n}^{m}(0)]^{2}},
\end{equation}
\end{widetext}
where $c_{mn}=(2-\delta _{0m})n(n+1)(2n+1)(n-m)!/(n+m)!$ and contributions from both the free-space field and the scattered field have been incorporated. From the definition, the spectral density characterizing the coupling strength between QEs and LSPs can be calculated as
\begin{widetext}
\begin{equation}
J_{lj}(\omega )=\frac{3\gamma _{0}\omega ^{3}\sqrt{\varepsilon _{\text{d}}}}{4\pi \omega _{0}^{3}}\text{Re}\left[\sum_{n=1}^{\infty }\sum_{m=0}^{n}c_{mn}\cos [\frac{2\pi m(l-j)}{N}]\frac{h_{n}^{(1)}(k_{1}r)[j_{n}(k_{1}r)+\mathcal{R}^{V}h_{n}^{(1)}(k_{1}r)]}{[k_{1}r/P_{n}^{m}(0)]^{2}}\right]. \label{spec}
\end{equation}
\end{widetext}
Defining $J_{lj}(\omega)\equiv J_{|l-j|}(\omega) $, we can verify that the spectral densities are periodic with $J_{l}(\omega)=J_{N-l}(\omega)$.

When the radius of the MNP is very small compared to the wavelength, i.e., $|k_{2}R|\ll 1,|k_{1}R|\ll 1,$, the Green's function can be further simplified. Substituting the limits
\begin{eqnarray}
\lim_{\rho\rightarrow 0}j_{n}(\rho ) &= &\frac{\rho ^{n}}{(2n+1)!!}, \\
\lim_{\rho\rightarrow 0}\partial _{\rho }[\rho j_{n}(\rho )] &= &\frac{(n+1)\rho ^{n}}{(2n+1)!!}, \\
\lim_{\rho\rightarrow 0}h_{n}^{(1)}(\rho ) &= &-i\frac{(2n-1)!!}{\rho ^{n+1}}, \\
\lim_{\rho\rightarrow 0}\partial _{\rho }[\rho h_{n}^{(1)}(\rho )] &=&\frac{i n(2n-1)!!}{\rho ^{n+1}},
\end{eqnarray}
into Eq. \eqref{coeff}, we can readily have $\mathcal{R}^{H}=0$ and $\mathcal{R}^{V}=\sum_{n=1}^{\infty }\mathcal{R}_{n}^{V}$, with
\begin{equation}\mathcal{R}_{n}^{V}=\frac{-i(k_{1}R)^{2n+1}(n+1)}{(2n+1)!!(2n-1)!!}\frac{
\varepsilon _\text{d}-\varepsilon _\text{m}(\omega )}{n\varepsilon _\text{m}(\omega
)+(n+1)\varepsilon _\text{d}}.\end{equation} Then the scattered Green's function can be decomposed into
\begin{equation}
\mathbf{G}^{\text{R}}(\mathbf{r},\mathbf{r}^{\prime},\omega )=\sum_{n=1}^{\infty }\mathbf{G}^{\text{R}}_{n}(\mathbf{r},\mathbf{r}^{\prime},\omega ),
\end{equation}
with the scattering coefficient in Eq. \eqref{sca} replaced by $\mathcal{R}_{n}^{V}$. The poles of $\mathcal{R}_{n}^{V}$ determines the resonance frequency of the LSPs. In this manner, the LSPs are expressed as a series of resonant modes labeled by $n$ with eigenfrequency determined by
\begin{equation}n\varepsilon _\text{m}(\omega_{n} )+(n+1)\varepsilon _\text{d}=0,
\end{equation} from which the contributions of the different resonant modes of LSPs to the light-matter interaction can be studied. In the low-frequency condition, the resonant frequencies can be determined by $\text{Re}[\varepsilon _\text{m}(\omega
_{n})]=-(n+1)\varepsilon _\text{d}/n$ due to $\textrm{Re}[\varepsilon _{m}(\omega )]\gg \textrm{Im}[\varepsilon _{m}(\omega )]$ \cite{PhysRevB.6.4370}. The first resonant mode is called the dipole mode and the second one is the quadrupole mode \cite{PhysRevLett.112.253601}. Using the parameters in our system, we can calculated the frequencies of the dipole and quadrupole modes
$\omega_{1}=3.77$ eV and $\omega_2=3.94$ eV.

\section{Derivation of the evolution equations}\label{appen-evolution}
In this appendix, we give the derivation of Eq. (2). In the single-excitation subspace, the time-evolved state can be expanded as $
|\Psi(t)\rangle=[\sum _lc_{l}(t)\hat{\sigma}_l^\dag+\int d^{3}\mathbf{r}\int d\omega d_{\mathbf{r},\omega}(t) \hat{\mathbf{f}}^\dag(\mathbf{r},\omega)]|G;\{0_{\omega}\}\rangle$. According to the Schr\"{o}dinger equation $i\hbar|\dot{\Psi} (t)\rangle =\hat{H}|\Psi (t)\rangle$, we have
\begin{widetext}
\begin{eqnarray}
\dot{c}_{l}(t) =-i\omega _{l}c_{l}(t)-\int d\omega \int d^{3}\mathbf{r}\frac{c^{-2}\omega ^{2}}{ \sqrt{\pi \varepsilon _{0}\hbar }}\sqrt{\text{Im}[\varepsilon _{m}(\omega )]}\mu _{l\hat{j}}G_{\hat{j}\hat{i}}(\mathbf{r}_{l},\mathbf{r},\omega )d_{\mathbf{r},\omega }(t), \label{cl} \\
\dot{d}_{\mathbf{r},\omega }(t) =-i\omega d_{\mathbf{r},\omega
}(t)+\sum_{l}\frac{c^{-2}\omega ^{2}}{ \sqrt{\pi \varepsilon _{0}\hbar
}}\sqrt{\text{Im}[\varepsilon _{m}(\omega )]}\mu _{l\hat{k}}^{\ast }G_{\hat{k}\hat{i}}^{\ast }(\mathbf{r}_{l},\mathbf{r},\omega )c_{l}(t), \label{cd}
\end{eqnarray}
\end{widetext}
with $l,j=0,\ldots,N-1$ and $\hat{i},\hat{j},\hat{k}=x,y,z$. Using $d_{\mathbf{r},\omega }(0)=0$, Eq. \eqref{cd} can be formally solved as
\begin{eqnarray}
d_{\mathbf{r},\omega }(t)&=&\sum_{l}\int_{0}^{t}d\tau e^{-i\omega (t-\tau )}\frac{c^{-2}\omega ^{2}}{\sqrt{\pi \varepsilon _{0}\hbar }}\sqrt{\text{Im}[\varepsilon _{m}(\omega )]}\nonumber \\ &&\times \mu _{l\hat{k}}^{\ast }G_{\hat{k}\hat{i}}^{\ast }(\mathbf{r}_{l},\mathbf{r},\omega )c_{l}(\tau ).
\end{eqnarray}
Substituting this into Eq. \eqref{cl}, we have
\begin{equation}
\dot{c}_{l}(t)+i\omega _{l}c_{l}(t)+\sum_{j}\int_{0}^{t}d\tau
\int_{0}^{\infty }d\omega e^{-i\omega (t-\tau )}J_{lj}(\omega )c_{j}(\tau )=0,
\end{equation}
with $J_{lj }(\omega )=\omega ^{2}\pmb{\mu}_{l}\cdot\textrm{Im}[ \mathbf{G}(\mathbf{r}_{l}, \mathbf{r}_{j},\omega )]\cdot \pmb{\mu}_{j}^{\ast }/(\pi\hbar \varepsilon_{0}c^{2})$ being the spectral density. We have used $\int d^3\mathbf{s}\frac{\omega ^{2}}{c^{2}}$Im$[\varepsilon _{m}(\omega )]\mathbf{G}(\mathbf{r},\mathbf{s},\omega )\mathbf{G}^{\ast }(\mathbf{r}^{\prime },\mathbf{s},\omega )=\textrm{Im}[\mathbf{G}(\mathbf{r},\mathbf{r}^{\prime },\omega )]$ \cite{PhysRevA.62.053804}.

For simplify, we choose the QEs having identical frequency $\omega_l=\omega_{0}$. Introducing a column vector $\mathbf{c}(t)=(c_{0}(t),c_{1}(t),\ldots ,c_{N-1}(t))^{T}$ and a spectral density matrix ${\bf J}(\omega)=J_{lj}(\omega)$, we obtain the evolution equation as Eq. (2).

\section{Eigenenergies of the system}\label{appen-eigenenergy}
In this appendix, we give the derivation of the energy spectrum of the whole system in the single-excitation subspace and the proof that they are exactly the same as the poles in the evolution equation under the Laplace transform in the main text.

The eigenstate $|\Phi\rangle $ of the QE-LSP system in the single-excitation subspace can be expanded as $|\Phi \rangle =[\sum_{l=0}^{N-1}c_{l}\hat{\sigma}_{l}^{\dag }+\int
d^{3}\mathbf{r}\int d\omega d_{\mathbf{r},\omega }\hat{\mathbf{f}}^{\dag }(\mathbf{r},\omega )]|G;\{0_{\omega }\}\rangle $. According to the stationary Schr\"{o}dinger equation $\hat{H}|\Phi\rangle =E|\Phi\rangle $, with $E$ the eigenenergy, we have
\begin{widetext}
\begin{eqnarray}
Ec_{l} &=&\hbar \omega _{0}c_{l}-i\hbar \int d\omega \int d^{3}\mathbf{r}%
^{\prime }\frac{c^{-2}\omega ^{2}}{ \sqrt{\pi \varepsilon _{0}\hbar }}%
\sqrt{\text{Im}[\varepsilon _{m}(\omega )]}\mu _{l\hat{j}}G_{\hat{j}\hat{i}}(%
\mathbf{r}_{l},\mathbf{r}^{\prime },\omega )d_{\mathbf{r},\omega},  \label{ca}
\\
Ed_{\mathbf{r},\omega} &=&\hbar \omega d_{\mathbf{r},\omega }+i\hbar
\sum_{j=0}^{N-1}\frac{c^{-2}\omega ^{2}}{\sqrt{\pi \varepsilon
_{0}\hbar }}\sqrt{\text{Im}[\varepsilon _{m}(\omega )]} \mu _{j\hat{k}}^{\ast }G_{\hat{k}\hat{i}}^{\ast }(\mathbf{r}_{l},\mathbf{r},\omega
)c_{l},
\end{eqnarray}\end{widetext}
with $l,j=0,\ldots,N-1$ and $\hat{i},\hat{j},\hat{k}=x,y,z$. Solving $d_{\mathbf{r},\omega}$ and substituting it into Eq. \eqref{ca}, it is easy to obtain
\begin{equation}
(E-\hbar \omega _{0})c_{l}-\hbar ^{2}\sum_{j =0}^{N-1}\int d\omega \frac{J_{lj}(\omega )}{E-\hbar \omega }c_{j}=0,\label{stadff}
\end{equation}
or
\begin{equation}
(E-\hbar \omega _{0})\mathbf{c}-\hbar ^{2}\int \frac{\mathbf{J}(\omega )d\omega }{E-\hbar\omega }\mathbf{c}=0, \label{e2}
\end{equation}
expressed in a matrix form. Using the Jordan decomposition of $\mathbf{J}(\omega )=\mathbf{V}\mathbf{D}(\omega )\mathbf{V}^{-1}$, with $\mathbf{V}$ and $\mathbf{D}(\omega )=\text{diag}[D_{0}(\omega),\ldots,D_{N-1}(\omega)]$ its similarity matrix and Jordan canonical form, Eq. \eqref{e2} can be expressed as
\begin{equation}
\lbrack E-\hbar \omega _{0}-\hbar ^{2}\int \frac{\mathbf{D}(\omega ) }{E-\hbar \omega }d\omega]\mathbf{\bar{c}}=0,
\end{equation}
where $\mathbf{\bar{c}}=\mathbf{V}^{-1}\mathbf{c}$. The equations have nontrivial solutions if and only if the determinant of the coefficient matrix is zero. Therefore, the eigenvalues of the QE-LSP system in the single-excitation subspace are determined by
\begin{equation}\label{e3}
E=\hbar \omega _{0}+\hbar ^{2}\int \frac{D_{l}(\omega )}{E-\hbar \omega }d\omega .
\end{equation}
Equation \eqref{e3} takes the same form as the equation to determine the bound state obtained in the main text. This clearly demonstrates that the dynamics of QEs essentially depends on the energy-spectrum character of the whole QE-LSP system.

\section{Solution in the steady state}\label{appen-solution}
Suppose $M$ bound states form outside the continuous energy band. Then, according to the completeness of the eigenstates, the time evolution of $|\Psi(0)\rangle=\hat{\sigma}_0^\dag|G;\{0_\omega\}\rangle$ can be expanded as
\begin{equation}
|\Psi(t)\rangle=\sum_{\alpha=1}^M x^b_\alpha e^{-i\varpi^b_\alpha t}|\Phi^b_\alpha\rangle+\sum_{E\in \text{CB}}x_E e^{-i E t/\hbar}|\Phi_E\rangle,
\end{equation}
where $x^b_\alpha=\langle \Phi_\alpha^b|\Psi(0)\rangle$ and $x_E=\langle \Phi_E|\Psi(0)\rangle$. The first term is contributed by the bound eigenstates and the second one is from the continuous-band eigenstates. Its overlap with the initial state reads
\begin{equation}
\langle\Psi(0)|\Psi(t)\rangle=\sum_{\alpha=1}^M |x^b_\alpha|^2 e^{-i\varpi^b_\alpha t}+\sum_{E\in \text{CB}}|x_E|^2 e^{-i E t/\hbar}.
\end{equation}
The initial-state fidelity defined as $P(t)=|\langle\Psi(0)|\Psi(t)\rangle|^2$ reads
\begin{eqnarray}
P(t)&=&\Big|\sum_{\alpha=1}^M |x^b_\alpha|^2 e^{-i\varpi^b_\alpha t}\Big|^2+\Big|\sum_{E\in \text{CB}}|x_E|^2 e^{-i E t/\hbar}\Big|^2\nonumber\\
&&+2\sum_{\alpha=1}^M\sum_{E\in \text{CB}}|x^b_\alpha|^2 |x_E|^2\cos(\varpi^b_\alpha-{E\over\hbar})t.
\end{eqnarray}
Both the second and the third term contain the oscillating frequencies $E/\hbar$, which are continuously summed in the continuous energy band. Such terms in the continuous energy band tend to vanish due to the out-of-phase interference of the different components in the long-time limit. Thus only the isolated bound states survive in the long-time limit. For the $N=2$ case, at most two bound states can be formed. Therefore, we have
\begin{equation}
\lim_{t\rightarrow \infty}P(t)=\left\{ \begin{aligned}
         &0,\hspace{2.5cm}~~M=0\\
         &|x^b|^4,\hspace{1.9cm}~~~M=1 \\
         &|x^b_1|^4+|x^b_2|^4+F(t),~M=2,
                          \end{aligned} \right. \label{stdv}
\end{equation}with $F(t)=2|x^b_1|^2|x^b_2|^2\cos(\varpi^b_1-\varpi^b_0)t$.

To determine $x^b_\alpha$, we solve the stationary Schr\"{o}dinger equation $\hat{H}|\Phi\rangle=E|\Phi\rangle$. From Eqs. (\ref{ca})$-$(\ref{stadff}) we have
\begin{eqnarray}
(E-\hbar\omega_0)c_0=\int_0^\infty d\omega{\hbar^2[J_0(\omega)c_0+J_1(\omega)c_1]\over E-\hbar\omega},\label{x0b}\\
(E-\hbar\omega_0)c_1=\int_0^\infty d\omega{\hbar^2[J_1(\omega)c_0+J_0(\omega)c_1]\over E-\hbar\omega},\label{x1b}\\
(E-\hbar \omega) d_{\mathbf{r},\omega }=i\hbar \sum_{l}\frac{c^{-2}\omega ^{2}}{ \sqrt{\pi \varepsilon _{0}\hbar }}\sqrt{\text{Im}[\varepsilon _{m}(\omega )]}\nonumber\\
\times\mu _{l\hat{k}}^{\ast }G_{\hat{k}\hat{\imath}}^{\ast }(\mathbf{r}_{l},\mathbf{r},\omega )c_{l}. \label{y}
\end{eqnarray}
Equation \eqref{y} leads to
\begin{equation}
\int d^{3}\mathbf{r}|d_{\mathbf{r},\omega }|^{2} =\frac{\hbar ^{2}[J_{0}(\omega)(|c_{0}|^{2}+|c_{1}|^{2})+J_{1}(\omega )(c_{1}c_{0}+c_{1}c_{0})]}{(E-\hbar \omega )^{2}}.\label{y2}
\end{equation}
Equations (\ref{x0b}) and (\ref{x1b}) have nontrivial solutions if and only if
\begin{eqnarray}
y_\pm(E)\equiv\hbar\omega_0+\int_0^\infty d\omega {\hbar^2[J_0(\omega)\pm J_1(\omega)]\over E-\hbar\omega}=E.~~\label{Engd}
\end{eqnarray}If $y_\pm(0)<0$, two bound states with the eigenenergies $E^b=\hbar\varpi^b_+$ and $\hbar\varpi^b_-$  determined by Eq. \eqref{Engd} can be formed in the band-gap area. Focusing on these bound states, we calculate their corresponding excited-state populations $|c_{0,\pm}^b|^2$ in the first QE. Substituting Eqs. (\ref{x0b}), (\ref{x1b}), and (\ref{y2}) into the normalization condition $\sum_{l=0}^{1}|c_{l}|^{2}+\int d^{3}\mathbf{r}\int_{0}^{\infty} d\omega|d_{\mathbf{r}, \omega }|^{2}=1$ and repeatedly using $E^b-\hbar\omega_0-\int_0^\infty d\omega{\hbar^2J_0(\omega)\over E^b-\hbar\omega}=\pm \int_0^\infty d\omega{\hbar^2J_1(\omega)\over E^b-\hbar\omega}$ obtained from Eq. (\ref{Engd}) for the bound states, we obtain
\begin{equation}
|c^b_{0,\pm}|^{2}=\frac{1}{2}\Big[1+\int_{0}^{\infty} d\omega \frac{\hbar ^{2}[J_{0}(\omega)\pm J_{1}(\omega)]}{(E^{b}-\hbar \omega )^{2}}\Big]^{-1}.
\end{equation}
It can be verified that $x_{1,2}^{b}=c_{0,\pm}^{b\ast}$. According to the forms of $Z_l$ obtained by the Laplace transform in the main text, we can readily see that $|x_{1,2}^{b}|^{2}={Z_{0,1}\over 2}$.

The above process gives the analytical proof to Eq. (5) from the bound states. From this proof we can clearly see the distinguished role of the formed bound states in lossless steady-state behaviors. Such suppression to the decay in the lossy medium is guaranteed by the characters of the bound states as stationary states with isolated eigenenergies of the whole system.

\section{Exact Dynamics for $N=4$}\label{appen-4QEs}
\begin{figure}[tbp]
\centering
\includegraphics[width=\columnwidth]{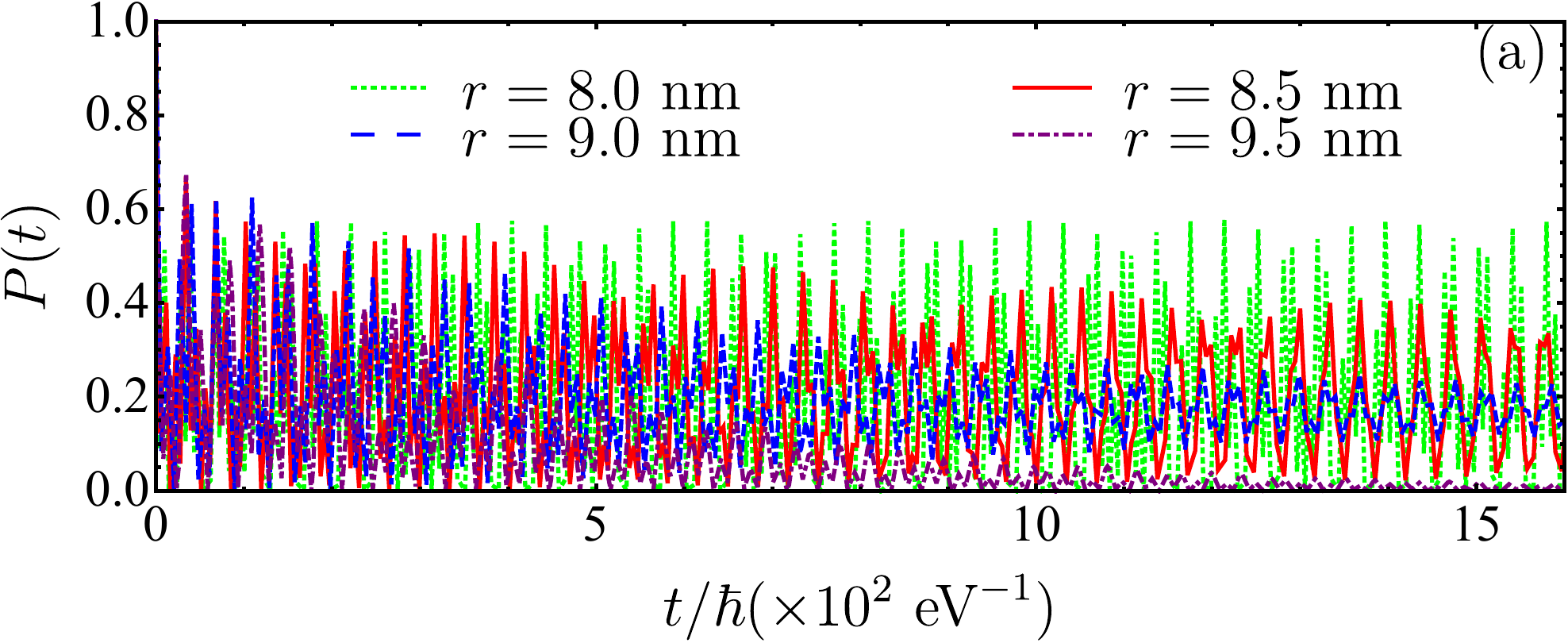}\\
\includegraphics[width=\columnwidth]{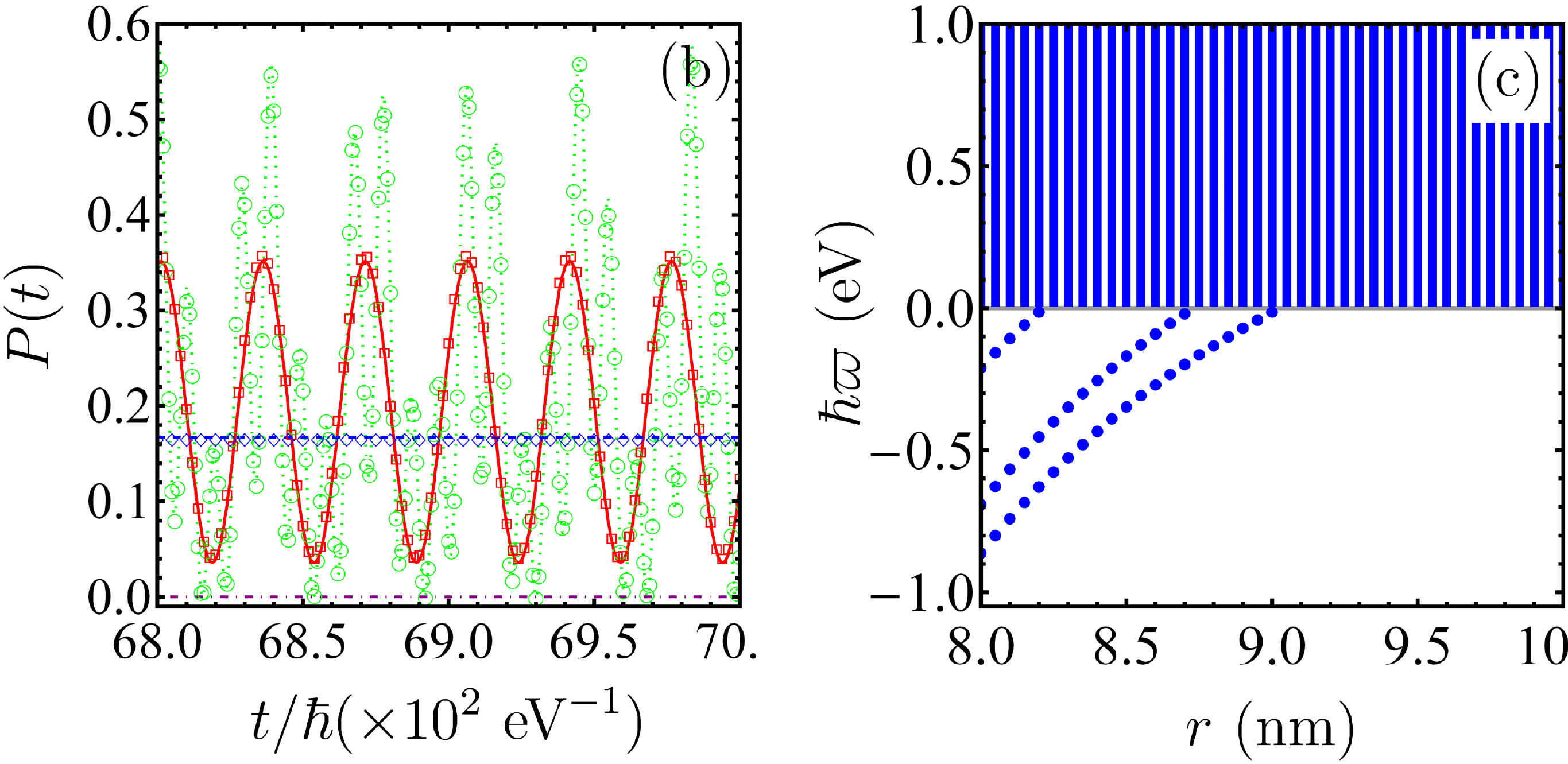}\\
\caption{(a) Evolution dynamics of $P(t)$ in different $r$ obtained by numerically solving Eq. \eqref{evolution-tr}. (b) Details of the dynamics in long-time limit. The circles, squares, and diamonds denote the long-time values of $P(t)$ obtained from the bound-state analysis, which correspond with the numerical results. (c) Energy spectrum of the whole system. The parameters are the same as in Fig. 2, but with $N=4$.} \label{Fig4}
\end{figure}

The spectral density matrix for $N=4$ reads
\begin{equation}
\mathbf{J}(\omega )=\left[
\begin{array}{cccc}
J_{0}(\omega ) & J_{1}(\omega ) & J_{2}(\omega ) & J_{1}(\omega ) \\
J_{1}(\omega ) & J_{0}(\omega ) & J_{1}(\omega ) & J_{2}(\omega ) \\
J_{2}(\omega ) & J_{1}(\omega ) & J_{0}(\omega ) & J_{1}(\omega ) \\
J_{1}(\omega ) & J_{2}(\omega ) & J_{1}(\omega ) & J_{0}(\omega )%
\end{array}%
\right],
\end{equation}
where the periodic condition $J_{l}(\omega)=J_{N-l}(\omega)$ has been used. As a symmetric and circulant matrix, $\mathbf{J}(\omega )=\mathbf{V}\mathbf{D}(\omega )\mathbf{V}^{-1}$, where $\mathbf{D}(\omega )={\text{diag}}[ J_{0}(\omega )+2J_{1}(\omega
)+J_{2}(\omega ),J_{0}(\omega )-J_{2}(\omega ),J_{0}(\omega )-2J_{1}(\omega
)+J_{2}(\omega ),J_{0}(\omega )-J_{2}(\omega )]$ and $\mathbf{V}=\frac{1}{2}\left[
\begin{array}{cccc}
1 & 1 & 1 & 1 \\
1 & i & -1 & -i \\
1 & -1 & 1 & -1 \\
1 & -i & -1 & i%
\end{array}%
\right] $ \cite{CIT-006}. Note that the eigenvalues are degenerate with $D_{1}(\omega)=D_{3}(\omega)$. The canonical transform $\mathbf{\bar{c}}(t)=\mathbf{V}^{-1}\mathbf{c}(t)$ can convert the integro-differential equation into
\begin{equation}
\mathbf{\dot{\bar{c}}}(t)+i\omega _{0}\mathbf{\bar{c}}(t)+
\int_{0}^{t}d\tau\int d\omega e^{-i\omega (t-\tau )}\mathbf{D}(\omega )\mathbf{\bar{c}}(\tau)=0, \label{evolution-tr}
\end{equation}
with \begin{equation}
\mathbf{\bar{c}}(t)=\frac{1}{2}\left[
\begin{array}{c}
c_{0}(t)+2c_{1}(t)+c_{2}(t) \\
c_{0}(t)-c_{2}(t) \\
c_{0}(t)-2c_{1}(t)+c_{2}(t) \\
0%
\end{array}%
\right]
\end{equation} and $\mathbf{\bar{c}}(0)=\frac{1}{2}\left[ 1,1,1,0\right] ^{T}$ under the initial condition $|\Psi (0)\rangle =\hat{\sigma}_0^{\dag}|G;\{0_\omega\}\rangle$, where $c_1(t)=c_3(t)$ has been used. Equation \eqref{evolution-tr} is analytically solvable by the Laplace transform. As shown in the main text, its solution in the long-time limit reads
\begin{equation}
\lim_{t\rightarrow \infty }\bar{c}_{l}(t)=\left\{ \begin{aligned}
         &(Ze^{-i\varpi_l^{b}t})\bar{c}_l(0),~y_l(0)<0 \\
         &0,\hspace{1.9cm}~y_l(0)>0.        \end{aligned} \right.
\end{equation}
 This clearly shows that the dynamics of the system in the long-time limit is determined by the formation of a bound of the whole system. It is not easy to find that the initial-state fidelity equals $P(t)=|c_0(t)|^2$, with $c_{0}(t)=\frac{1}{4}[\bar{c}_{0}(t)+2\bar{c}_{1}(t)+\bar{c}_{2}(t)]$.

Figure \ref{Fig4}(a) plots the evolution of $P(t)$ in different $r$. The different behaviors, i.e., complete decay, population trapping, and persistent oscillation, are present depending on the value of $r$. Details on the long-time behaviors are shown in Fig. \ref{Fig4}(b). Such phenomena are associated with the formation of the bound state of the QE-LSP system. Figure \ref{Fig4} (c) shows the energy spectrum of the whole system. If no bound state is formed, then $P(t)$ tends to zero, which characterizes the complete decoherence. If one bound state is formed, then $P(t)$ tends to a finite value, which describes the population trapping. If two or more bound states are formed, then $P(t)$ tends to the Rabi-like persistent oscillations in the long-time limit. Such behaviors coincide with our analytical analysis.

\bibliography{LSPs}
\end{document}